\documentclass[aps,amsmath,prl,reprint,showpacs,preprintnumbers,twocolumn]{revtex4}
\usepackage{amsfonts}
\usepackage{graphicx}
\usepackage{dcolumn}
\usepackage{pdfsync}
\usepackage{multirow}
\usepackage{subfigure}
\usepackage{palatino}
\usepackage{graphicx}
\usepackage{epsfig}
\usepackage{ifpdf}
\ifpdf \DeclareGraphicsExtensions{.pdf,.png,.jpg,.mps}
 \else
\DeclareGraphicsExtensions{.eps,.pdf} \fi

\begin{document}
\title{A New Limit on Possible Long-Range Parity-odd Interactions of the Neutron from Neutron Spin Rotation in Liquid $^{4}He$}
\author{H.Yan}
\affiliation{Indiana University, Bloomington, Indiana 47408, USA}
\affiliation{Center for Exploration of Energy and Matter, Indiana University, Bloomington, IN 47408}
\author{W. M. Snow}
\email[Corresponding author: ]{wsnow@indiana.edu}
\affiliation{Indiana University, Bloomington, Indiana 47408, USA}
\affiliation{Center for Exploration of Energy and Matter, Indiana University, Bloomington, IN 47408}

\date{\today}
\begin{abstract}
Various theories beyond the Standard Model predict new particles with masses in the sub-eV range with very weak couplings to ordinary matter. A parity-odd interaction between polarized nucleons and unpolarized matter proportional to $g_{V}g_{A}{\vec{s}} \cdot {\vec{p}}$ is one such possibility, where ${\vec{s}}$ and ${\vec{p}}$ are the spin and the momentum of the polarized nucleon, and $g_{V}$ and $g_{A}$ are the vector and axial vector couplings of an interaction induced by the exchange of a new light vector boson. We report a new experimental upper bound on such possible long-range parity-odd interactions of the neutron with nucleons and electrons from a recent search for parity violation in neutron spin rotation in liquid $^{4}He$. Our constraint on the product of vector and axial vector couplings of a possible new light vector boson is  $g_{V}g_{A}^{n} \leq 10^{-32}$ for an interaction range of 1m. This upper bound is more than seven orders of magnitude more stringent than the existing laboratory constraints for interaction ranges below 1m, corresponding to a broad range of vector boson masses above $10^{-6}$ eV. More sensitive searches for a $g_{V}g_{A}^{n}$ coupling could be performed using neutron spin rotation measurements in heavy nuclei or through analysis of experiments conducted to search for nucleon-nucleon weak interactions and nuclear anapole moments.
\end{abstract}
\pacs{13.88.+e, 13.75.Cs, 14.20.Dh, 14.70.Pw}
\maketitle

\section{Introduction}

The possible existence of new interactions of nature with ranges of mesoscopic scale (millimeters to microns) and very weak couplings to matter has been mentioned occasionally in the past~\cite{Leitner,Hill} and has recently begun to attract more scientific attention. Particles which might transmit such interactions are starting to be referred to generically as WISPs (Weakly-Interacting sub-eV Particles)~\cite{Jae10} in recent theoretical literature. Many theories beyond the Standard Model possess extended symmetries which, when broken at a high energy scale, lead to weakly-coupled light particles with relatively long-range interactions, such as axions, familons, and Majorons~\cite{PDG12}.  Several theoretical attempts to explain dark matter and dark energy also produce new weakly-coupled long-range interactions. The fact that the dark energy density of order (1 meV)${^4}$  corresponds to a length scale of $~100$ $\mu$m  also encourages searches for new phenomena around this scale~\cite{Ade09}.

Experimental constraints on possible new interactions of mesoscopic range which depend on the spin of one or both of the particles are much less stringent than those for spin-independent interactions~\cite{Antoniadis11}. A general classification of interactions between nonrelativistic fermions assuming only rotational invariance~\cite{Dob06} uncovered 16 different operator structures involving the spins, momenta, interaction range, and various possible couplings of the particles. Of these sixteen interactions, one is spin-independent, six involve the spin of one of the particles, and the remaining nine involve both particle spins. Ten of these 16 possible interactions depend on the relative momenta of the particles.

In particular, there are very few laboratory constraints on possible new interactions of mesoscopic range which depend on {\it both} the spin {\it and} the relative momentum, since the polarized electrons or nucleons in most experiments employing macroscopic amounts of polarized matter typically possess $\langle\vec{p}\rangle=0$ in the lab frame.  Such interactions can be generated by a light vector boson $X_{\mu}$ coupling to a fermion $\psi$ with an interaction of the form  $\mathcal{L}_{I}=\bar{\psi}(g_{V}\gamma^{\mu}+g_{A}\gamma^{\mu}\gamma_{5})\psi X_{\mu}$ where $g_{V}$ and $g_{A}$ are the vector and axial couplings. In the nonrelativistic limit this interaction gives rise to two interaction potentials of interest depending on both the spin and the relative momentum~\cite{Pie11}: one proportional to $g_{A}^{2}\vec{\sigma}\cdot(\vec{v}\times\hat{r})$ and another proportional to $g_{V}g_{A}\vec{\sigma}\cdot\vec{v}$.
As noted above many theories beyond the Standard Model can give rise to such interactions. For example, spontaneous symmetry breaking in the Standard Model with two or more Higgs doublets with one doublet responsible for generating the up quark masses and the other generating the down quark masses can possess an extra U(1) symmetry generator distinct from those which generate $B$, $L$, and weak hypercharge $Y$. The most general U(1) generator in this case is some linear combination $F=aB + bL +cY + dF_{ax}$ of $B$, $L$, $Y$, and an extra axial U(1) generator $F_{ax}$ acting on quark and lepton fields, with the values of the constants $a,b,c,d$ depending on the details of the theory. The new vector boson associated with this axial generator can give rise to $\mathcal{L}_{I}$ above~\cite{Fayet:1990}.

Neutrons have recently been used with success to tightly constrain possible weakly coupled interactions of mesoscopic range~\cite{Dubbers11}. A polarized beam of slow neutrons can have a long mean free path in matter and is a good choice for such an experimental search~\cite{Nico05b}. Piegsa and Pignol~\cite{Pie12} recently reported improved constraints on the product of axial vector couplings $g_{A}^{2}$ in this interaction. Polarized slow neutrons which pass near the surface of a plane of unpolarized bulk material in the presence of such an interaction experience a phase shift which can be sought using Ramsey's well-known technique of separated oscillating fields~\cite{Ramsey:1950}.

In this paper we report a new constraint on $g_{V}g_{A}^{n}$. In the nonrelativistic limit $\mathcal{L}_{I}$ gives rise to the potential:
  \begin{equation}\label{eqn1}
  V(r)=\frac{g_{V}g_{A}}{2\pi}\frac{\exp{(-r/\lambda)}}{r}\vec{\sigma}\cdot\vec{v}
  \end{equation}
where $\lambda=1/m_{X}$ is the interaction range, $m_{X}$ is the mass of the vector boson, $\vec{s}=\vec{\sigma}/2$ is the spin of the polarized particle, and $r$ is the distance between the two interacting particles.To derive our constraint we take advantage of the fact that a term in the neutron interaction potential proportional to $\vec{\sigma}\cdot\vec{v}$ violates parity and therefore causes a rotation of the plane of polarization of a transversely polarized slow neutron beam as it moves through matter. This phenomenon is known as neutron optical activity in analogy with the well-known corresponding phenomenon of optical activity of transversely polarized light as it moves through a chiral medium.  The parity-violating (PV) interaction between the neutrons and the medium causes the amplitudes of the positive and negative neutron helicity states of polarized neutrons to accumulate different phases as they pass through the medium. The difference $\phi_{PV}$ between the phase shifts of the two neutron helicities upon motion through the medium leads to a rotation of the neutron polarization vector about its momentum, which manifestly violates parity~\cite{Mic64}. Parity-odd neutron spin rotation has been observed in heavy nuclei~\cite{Forte80, Heckel82, Heckel84}. The rotation angle per unit length $d\phi_{PV}/dL$ of a neutron of wave vector $k_{n}$ in a medium of density $\rho$ is $d\phi_{PV}/dL=4\pi \rho f_{PV}/k_{n}$, where $f_{PV}$ is the forward limit of the parity-odd p-wave scattering amplitude. Because $f_{PV}$ is proportional to the parity-odd correlation $\vec{\sigma}_{n}\cdot \vec{k}_{n}$ with $\vec{\sigma}_{n}$ the neutron spin vector, $d\phi_{PV}/dL$ is constant as $k \to 0$ in the absence of resonances~\cite{Sto82}. For the spin-velocity interaction described by Eqn.(\ref{eqn1}),for an infinite large sample which the polarized neutron passing through, one can apply the Born approximation to derive the relation between $f_{PV}$ and the parameters of the potential, and the spin rotation angle per unit length can be expressed directly in terms of the vector and axial vector couplings,  the range of the interaction, and the number density:
\begin{equation}\label{eqn2}
\frac{d\phi_{PV}}{dL}=4g_{V}g_{A}\rho\lambda^{2}
\end{equation}

\begin{figure*}
\begin{center}
\includegraphics[scale=1.2]{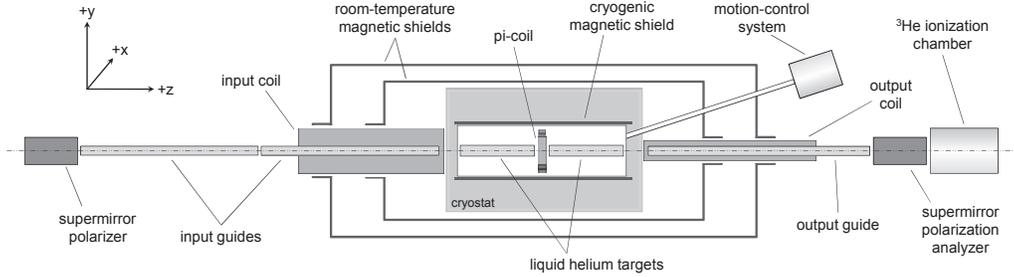}
\caption{Overview of the apparatus used to measure parity-odd neutron spin rotation in liquid helium.}
\label{fig:apparatus}
\end{center}
\end{figure*}



\section{Experimental Technique, Measurement, and Results}

An experiment to search for parity-odd neutron spin rotation in liquid $^{4}$He was performed at the NG-6 slow neutron beamline at the National Institute of Standards and Technology (NIST) Center for Neutron Research~\cite{Nico05}. The apparatus shown in Figure 1 must distinguish small PV rotations from rotations that arise from residual magnetic fields. $\phi_{PV}$ is isolated by alternately moving the liquid in front of and behind a neutron spin precession coil (called the $\pi$-coil in the figure) and measuring the change in the spin rotation angle using the neutron equivalent of a crossed polarizer/analyzer pair familiar from light optics.  Neutrons polarized along  ${\hat y}$ enter a central precession coil with an internal magnetic field along  ${\hat y}$ ( from the $\pi$-coil) which precesses any spin component along ${+\hat x}$ to  ${-\hat x}$. With the $\pi$-coil turned on, the contribution to the total rotation angle coming from parity violation in the liquid changes sign as the liquid is moved from the upstream target chamber to the downstream target chamber. To further suppress systematic uncertainties and noise, the beam and apparatus are split into right and left halves, and the targets are filled so that the liquid occupies the chamber downstream of the $\pi$-coil on one half of the beam and the chamber upstream of the $\pi$-coil on the other half of the beam. The PV components of the neutron spin rotation angle from the liquid target therefore possess opposite signs on each side of the beam, and the difference of the two rotation angles is insensitive to both static residual magnetic fields and any common-mode time-dependent magnetic field integrals seen by the neutrons along their trajectories in the target. The experiment, apparatus, and analysis of systematic errors has been described in detail elsewhere~\cite{Bas09, Sno11, Micherdzinska11, Swa10, Sno12}.

\begin{figure}[h]
\begin{center}
\includegraphics[scale=0.4]{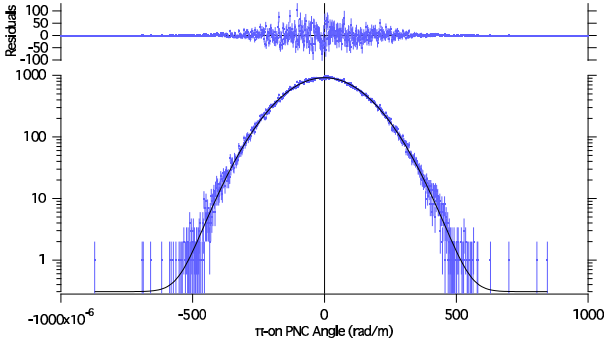}
\caption{Distribution of measured spin rotation angles per meter in liquid $^{4}$He. The solid lines are fits to a Gaussian distribution with a constant background.}
\label{fig:data}
\end{center}
\end{figure}

The measured upper bound on the parity-odd neutron spin rotation angle per unit length in liquid $^{4}He$ at a temperature of 4K based on the data shown in Figure 2 is $d\phi_{PV}/dL=+1.7\pm9.1(stat.)\pm1.4(sys)\times10^{-7}rad/m$. We can derive a limit on $g_{V}g_{A}^{n}$ directly from Eqn.(\ref{eqn2}) until the interaction range becomes comparable to the size of the target medium: in this regime we must perform a numerical integration to relate the spin rotation angle to the parameters of the potential for our experimental geometry. Each of the four internal target chambers holding the liquid helium in the experiment had dimensions 40 cm$\times$ 2.5 cm $\times$ 5 cm and the transversely polarized neutron beam of size 2 cm $\times$ 5 cm uniformly illuminates the targets. The 1 m target region length sets an upper bound on the interaction range we are sensitive to. Our constraint on the product $g_{V}g_{A}^{n}$, shown as the solid line in Figure 3, ranges from $g_{V}g_{A}^{n} \leq 10^{-32}$ at 1m to $g_{V}g_{A}^{n} \leq 10^{-22}$ at 1 micron. The corresponding range of vector boson masses varies from $\sim10^{-6}$ to $\sim10^{-1}$ eV. This upper bound is more than $7$ orders of magnitude more stringent at a distance of 1m than the best existing laboratory constraint,  which comes from an atomic physics measurement using a polarized $K$-$^{3}He$ co-magnetometer technique~\cite{Vas09}. The grey region in the figure shows the new combinations of coupling strength and range ruled out by this work down to a distance of 1 micron: at smaller distances the constraint follows Eqn.(\ref{eqn2}). Since neutron spin rotation involves the real part of the coherent forward scattering amplitude the vector coupling constrained in this experiment applies to an equal number of protons, neutrons, and electrons.

 \begin{figure}[h]
\centering
\includegraphics[scale=1.4]{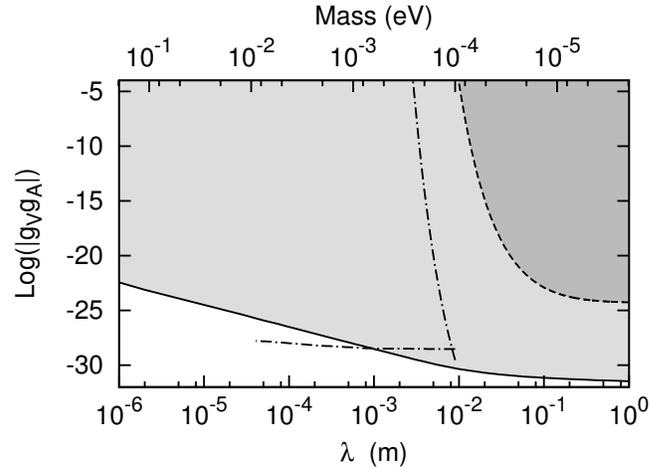}
 \caption{\small{Upper bounds on the product of couplings $g_{V}g_{A}^{n}$ for a possible long-range parity-odd interaction of the neutron with nucleons and electrons. The regions excluded by laboratory experiments for ranges between 1 micron and 1 meter are shown: constraints at shorter distances follow Eqn.(\ref{eqn2}) . The dashed line comes from~\cite{Vas09}. The light grey region above the solid line shows the new regions excluded by this work.}}
\label{fig:constraints}
\end{figure}


A background in this measurement can in principle come from quark-quark weak interactions present in the Standard Model, which induce weak interactions between nucleons that violate parity. It is not yet possible to calculate the Standard Model contribution to parity-odd neutron spin rotation in this system given our inability to deal with the strongly interacting limit of QCD, and indeed the weak interaction between nucleons remains one of the most poorly-understood aspects of low energy weak interaction physics. One can roughly estimate the expected size of NN weak interaction amplitudes relative to strong interaction amplitudes to be of order  $10^{-6}$ to $10^{-7}$ for energies far below the electroweak scale~\cite{Sto74}. The best existing estimate of $d\phi_{PV}/dL$ in n-$^{4}$He from Standard Model weak interactions was derived using existing measurements of nuclear parity violation in a model~\cite{Des98} which subsumes many poorly-understood short-range strong NN interaction effects in nuclei by expressing parity-odd amplitudes in terms of isoscalar ($X_{n}+X_{p}$) and isovector ($X_{n}-X_{p}$) one-body effective potentials. In this model n-$^{4}$He spin rotation is directly related to $X_{n}$. Existing measurements~\cite{Fluorine} and theoretical calculations~\cite{Haxton} for parity violation in $^{18}$F constrain $X_{n}-X_{p}$, and measurements in odd-proton systems such as p-$^{4}$He~\cite{Lang85, Lang86} and $^{19}$F~\cite{Els84} constrain $X_{p}$.  The resulting prediction for n-$^{4}$He spin rotation is $d\phi_{PV}/dL=(-6.5\pm2.2)\times 10^{-7}$ rad/m. Our experimental upper bound is larger than this estimate of the Standard Model background and we therefore ignore the unlikely possibility of a cancellation between this contribution and that from the term of interest in this work.


\section{Conclusion}

Slow neutron spin rotation is a very sensitive technique to search for possible exotic long-range neutron interactions which violate parity. By analyzing our recent upper bound on neutron spin rotation in liquid $^{4}He$~\cite{Sno11}, we derive an upper bound on the product of couplings $g_{V}g_{A}^{n}$ from any new long-range parity-odd interaction mediated by vector boson exchange. This constraint is more than 7 orders of magnitude more stringent than the current existing laboratory constraints over several decades of length scales below 1m.

It is interesting to consider how these constraints might be improved in a dedicated experiment. It is difficult to improve the constraint on $g_{V}g_{A}^{n}$
by repeating the helium spin rotation measurement with greater accuracy due to the Standard Model background discussed above expected from quark-quark weak interactions. Another Standard Model background from the parity-odd neutron-electron interaction also exists, but is suppressed compared to neutron-nucleon parity violation by a factor of $1-4 \sin^{2}\theta_{W} \approx 0.1$ and can be calculated to high accuracy. On the other hand this constraint from the isoscalar $^{4}$He nucleus is blind to possible vector couplings proportional to isospin which appear in some models. One could consider a measurement of neutron spin rotation using a target of much higher nucleon density with nonzero isospin and long mean free path. Better limits on $g_{V}g_{A}^{n}$ using polarized neutrons might also come from a measurement in progress of neutron-proton parity violation through a search for the parity-odd gamma asymmetry in polarized slow neutron capture on protons by the NPDGamma collaboration~\cite{Ger11}. To high accuracy this asymmetry is dominated by a single weak nucleon-nucleon amplitude involving pion exchange~\cite{LIU,SCHIAVILLA,DESPLANQUES01,HYUN01,HYUN05}, and both previous theoretical and experimental work along with recent calculations in lattice gauge theory~\cite{Wasem12} indicate that Standard Model parity violation might be suppressed in this system. Previous analysis of the comparison between  the measurement of the weak charge of the $^{133}$Cs atom and the Standard Model prediction~\cite{Bouchait:2005} placed tight constraints on $g_{V}^{n}g_{A}^{e}$. We expect that constraints on $g_{V}g_{A}^{n}$ could be derived from existing measurements of parity violation in atoms sensitive to the nuclear anapole moment, which comes from parity violating interactions between nucleons~\cite{Zel57, Woo97}.


This work was supported by the Department of Energy and by NSF grant PHY-1068712. We thank F. Piegsa and G. Pignol for communications. The work of WMS was supported in part by the Indiana University Center for Spacetime Symmetries.


\end{document}